\def\etc{{\it etc.}}

\def\ie{{\it i.e.}}

\def\~{{$\tilde{\phantom{a}}$}}

\documentclass [12pt] {article}
\usepackage{epsfig}
\usepackage{color}

\def\thebibliography#1{\section{References}\markboth
 {REFERENCES}{REFERENCES}\list
 {[\arabic{enumi}]}{\settowidth\labelwidth{[#1]}\leftmargin\labelwidth
 \advance\leftmargin\labelsep
 \usecounter{enumi}}
 \def\newblock{\hskip .11em plus .33em minus -.07em}
 \sloppy
 \sfcode`\.=1000\relax}
\def\upcite#1{\raise6pt\hbox{\scriptsize
\cite{#1}}}
  \def\lsim{\mathrel {\vcenter {\baselineskip 0pt \kern 0pt
    \hbox{$<$} \kern 0pt \hbox{$\sim$} }}}
    \def\gsim{\mathrel {\vcenter {\baselineskip 0pt \kern 0pt
    \hbox{$>$} \kern 0pt \hbox{$\sim$} }}}

\def\hline{\noalign{\hrule \vskip2pt}}

\def\|{\ifmmode\Vert\else \char`\|\fi}
\ifx\oldzeta\undefined                          
  \let\oldzeta=\zeta                            
  \def\zzeta{{\raise 2pt\hbox{$\oldzeta$}}}     
  \let\zeta=\zzeta                              
\fi

\ifx\oldchi\undefined                           
  \let\oldchi=\chi                              
  \def\cchi{{\raise 2pt\hbox{$\oldchi$}}}       
  \let\chi=\cchi                                
\fi

\def\frac#1#2{{#1 \over #2}}

\def\half{\ifinner {\scriptstyle {1 \over 2}}
   \else {1 \over 2} \fi}

\def\abs#1{\left\vert#1\right\vert}	

\def\simge{\mathrel{%
   \rlap{\raise 0.511ex \hbox{$>$}}{\lower 0.511ex \hbox{$\sim$}}}}
\def\simle{\mathrel{
   \rlap{\raise 0.511ex \hbox{$<$}}{\lower 0.511ex \hbox{$\sim$}}}}

\def\buildchar#1#2#3{{\null\!                   
   \mathop#1\limits^{#2}_{#3}                   
   \!\null}}                                    
\def\overcirc#1{\buildchar{#1}{\circ}{}}

\def\slashchar#1{\setbox0=\hbox{$#1$}           
   \dimen0=\wd0                                 
   \setbox1=\hbox{/} \dimen1=\wd1               
   \ifdim\dimen0>\dimen1                        
      \rlap{\hbox to \dimen0{\hfil/\hfil}}      
      #1                                        
   \else                                        
      \rlap{\hbox to \dimen1{\hfil$#1$\hfil}}   
      /                                         
   \fi}                                         %

\def\subrightarrow#1{
  \setbox0=\hbox{
    $\displaystyle\mathop{}
    \limits_{#1}$}
  \dimen0=\wd0
  \advance \dimen0 by .5em
  \mathrel{
    \mathop{\hbox to \dimen0{\rightarrowfill}}
       \limits_{#1}}}                           

\def\overlay#1#2{\ifmmode%
\setbox0=\hbox{$#1$}%
\setbox1=\hbox to\wd0{\hss$#2$\hss}\else%
\setbox0=\hbox{#1}%
\setbox1=\hbox to\wd0{\hss#2\hss}\fi%
#1\hskip-\wd0\box1 }

\def\pmb#1{\leavevmode\setbox0=\hbox{#1}%
\kern-.02em\copy0\kern-\wd0
\kern.04em\copy0\kern-\wd0
\kern-.02em\raise.04em\box0 }

\def\vereq#1#2{\lower3pt\vbox{\baselineskip1.5pt \lineskip1.5pt
\ialign{$\m@th#1\hfill##\hfil$\crcr#2\crcr\sim\crcr}}}

\def\tensor#1{\protect\@ontopof{#1}{\leftrightarrow}{1.15}\mathord{\box2}}
\def\overstar#1{\protect\@ontopof{#1}{\ast}{1.15}\mathord{\box2}}
\def\overdots#1{\protect\@ontopof{#1}{\cdots}{1.0}\mathord{\box2}}
\def\overcirc#1{\protect\@ontopof{#1}{\circ}{1.2}\mathord{\box2}}
\def\loarrow#1{\protect\@ontopof{#1}{\leftarrow}{1.15}\mathord{\box2}}
\def\roarrow#1{\protect\@ontopof{#1}{\rightarrow}{1.15}\mathord{\box2}}

\def\@ontopof#1#2#3{%
{\mathchoice
{\@@ontopof{#1}{#2}{#3}\displaystyle\scriptstyle}%
{\@@ontopof{#1}{#2}{#3}\textstyle\scriptstyle}%
{\@@ontopof{#1}{#2}{#3}\scriptstyle\scriptscriptstyle}%
{\@@ontopof{#1}{#2}{#3}\scriptscriptstyle\scriptscriptstyle}%
}%
}

\def\@@ontopof#1#2#3#4#5{%
\setbox0=\hbox{$#4#1$}%
\setbox1=\hbox{$#5#2$}%
\setbox2=\hbox{}\ht2=\ht0 \dp2=\dp0 %
\ifdim\wd0>\wd1 %
\setbox1=\hbox to\wd0{\hss\box1\hss}%
\mathord{\rlap{\raise#3\ht0\box1}\box0}%
\else   %
\setbox1=\hbox to.9\wd1{\hss\box1\hss}%
\setbox0=\hbox to\wd1{\hss$#4\relax#1$\hss}%
\mathord{\rlap{\copy0}\raise#3\ht0\box1}%
\fi
}%

\def\lambdabar{\protect\@lambdabar}
\def\@lambdabar{%
\relax
\bgroup
\def\@tempa{\hbox{\raise.73\ht0
\hbox to0pt{\kern.25\wd0\vrule width.5\wd0
height.1pt depth.1pt\hss}\box0}}%
\mathchoice{\setbox0\hbox{$\displaystyle\lambda$}\@tempa}%
{\setbox0\hbox{$\textstyle\lambda$}\@tempa}%
{\setbox0\hbox{$\scriptstyle\lambda$}\@tempa}%
{\setbox0\hbox{$\scriptscriptstyle\lambda$}\@tempa}%
\egroup
}

\def\corresponds{{\lower.2ex\hbox{=}}{\rm\kern-.75em^\triangle}}
\def\succsim{\succ\kern-.9em_\sim\kern.3em}
\def\precsim{\prec\kern-1em_\sim\kern.3em}
\def\slantfrac#1#2{\kern1em^{#1}\kern-.3em/\kern-.1em_{#2}}

\begin{document}

\begin{center}
{\Large\bf Bunching of Photons When Two Beams Pass Through
\\

\medskip
 a Beam Splitter}
\\

\medskip

Kirk T.~McDonald
\\
{\sl Joseph Henry Laboratories, Princeton University, Princeton, NJ 08544}
\\
Lijun J.~Wang
\\
{\sl NEC Research Institute, Inc., Princeton, NJ 08540}
\\
(Aug.\ 17, 2003)
\end{center}

\section{Problem}

Dirac has written  \cite{Dirac} ``Each photon then interferes only with itself.  
Interference between two different photons never occurs."  Indeed, a practical
definition is that ``classical'' optics consists of phenomena due to 
the interference of photons only with themselves.  However, photons obey Bose
statistics which implies a ``nonclassical'' tendency for them to ``bunch''.

For a simple example of nonclassical optical behavior, consider
two pulses containing $n_1$ and $n_2$ photons of a single frequency
that are simultaneously incident on two sides of a lossless, 50:50 beam 
splitter, as shown in the figure.
Deduce the probability that $N_1$ photons are observed in the direction
of beam 1, where $0 \leq N_1 \leq n_1 + n_2$ for a lossless splitter.

\medskip
\centerline{\includegraphics*[width=2in]{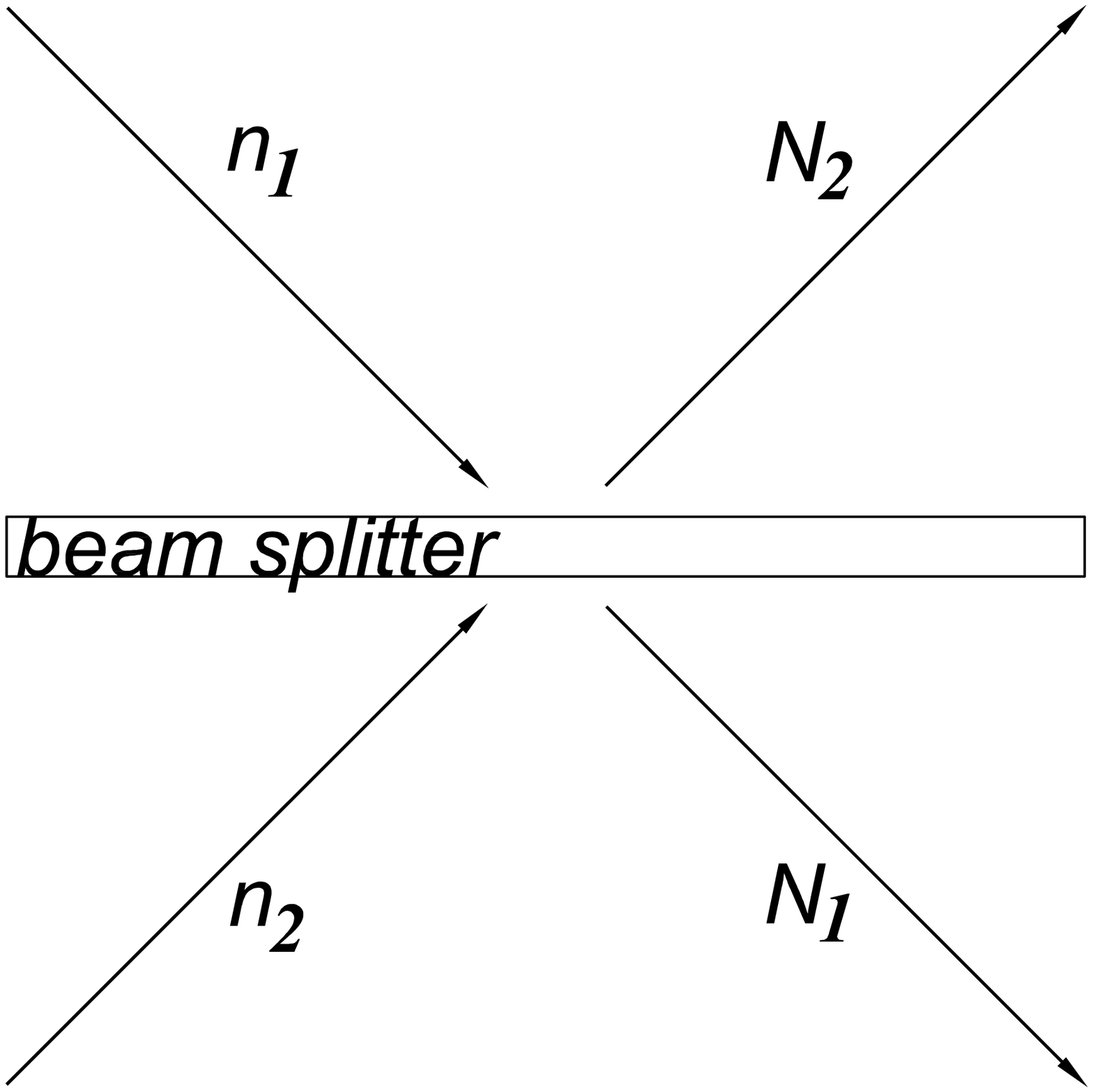}}

Hint: a relatively elementary argument can be given by recalling that the
phase of a reflected photon (\ie, of the reflected wave from a single input beam)
is $90^0$ different from that of a transmitted
photon \cite{set4}.  Consider first
the cases that $n_1$ or $n_2$ is zero.

\section{Solution}

An elegant solution can be given by noting the the creation and annihilation
operators relevant to a beam splitter obey an SU(2) symmetry \cite{Campos,Paris}.
Here, we give a more elementary solution, in the spirit of Feynman \cite{Feynman}.

Experimental demonstration of the case where $n_1 = n_2 = 1$ was first given
in \cite{Hong}, and the case of $n_1 = n_2 = 2$ has been studied in \cite{Ou}.

\subsection{A Single Input Beam}

We first consider the case of a single input beam with $n_1 > 0$.  Then, of course,
$n_2 = 0$.

In a classical view, the input beam would have energy $u_1 = n_1 \hbar \omega$,
where $\omega$ is the angular frequency of the photons.  Then, the effect of
the 50:50 beam splitter would be to create output beams of equal energies,
$U_1 = U_2 = u_1 / 2$.  In terms of photon numbers, the classical view would
imply that the only possibility for the output beams is $N_1 = N_2 = n_1 / 2$.

But in fact, the transmitted beam can contain any number $N_1$ of photons between
0 and $n_1$, while the reflected beam contains $N_2 - n_1 - N_1$ photons.

If the photons were distinguishable, we would assign a probability of
$(1 / 2)^{n_1}$ to each configuration of transmitted and reflected
photons in the 50:50 splitter.  But the photons are indistinguishable,
so that the probability that $N_1$ out of $n_1$ photons are transmitted
is larger than $(1 / 2)^{n_1}$ by the number of ways the $n_1$ photons
 can be arranged into a group of $N_1$ transmitted and $n_1 - N_1$
reflected photons with regard to their order, \ie, by the binomial coefficient,
\begin{equation}
C^{n_1}_{N_1} = {n_1! \over N_1! (n_1 - N_1)!}\, .
\label{s1}
\end{equation}
Thus, the probability $P(N_1,n_1 - N_1|n_1,0)$ that $N_1$ out of $n_1$ photons
(in a single input beam) are transmitted by the beam splitter is
\begin{equation}
P(N_1,n_1 - N_1|n_1,0) = C^{n_1}_{N_1}\left( {1 \over 2} \right)^{n_1}.
\label{s2}
\end{equation}

The result (\ref{s2}) is already very nonclassical, in that there is a
small, but nonzero probability that the entire input beam is transmitted,
or reflected.  However, in the limit of large $n_1$ the largest
probability is that the numbers of photons in the reflected and
transmitted beams are very nearly equal.  We confirm this by use of
Stirling's approximation for large $n$,
\begin{equation}
n! \approx e^{-n} n^n \sqrt{2 \pi n}.
\label{s3}
\end{equation}
For large $n$, and $k = (1 + \epsilon) n / 2$, we have
\begin{eqnarray}
C^n_k & \approx & {1 \over \sqrt{2 \pi n} \left( {k \over n} \right)^{k + 1/2}
\left( 1 - {k \over n} \right)^{n - k + 1/2} }
= {2^{n + 1} \over \sqrt{2 \pi n} (1 - \epsilon^2)^{(n + 1) / 2} 
\left( {1 + \epsilon \over 1 - \epsilon} \right)^{n \epsilon / 2}}
\nonumber \\
& \approx & {2^{n + 1} \over \sqrt{2 \pi n} (1 + n \epsilon^2 / 2) }\, .
\label{s4}
\end{eqnarray}
The probability of $k$ photons out of $n$ being transmitted drops to 1/2
the peak probability when $\epsilon \approx \sqrt{2 / n}$.  Hence, for
large $n$ the number distribution of photons in the transmitted (and reflected)
 beam is essentially a delta function centered at $n/2$, in agreement with the
classical view.

The most dramatic difference between the classical and quantum behavior of a 
single beam in a 50:50 beam splitter occurs when $n_1  = 2$,
\begin{equation}
P(0,2|2,0) = {1 \over 4}\, ,
\qquad
P(1,1|2,0) = {1 \over 2}\, ,
\qquad
P(2,0|2,0) = {1 \over 4}\, .
\label{s5}
\end{equation}

In the subsequent analysis we shall need to consider interference effects,
so we note that the magnitude of the
probability amplitude that $k$ out of $n$ photons 
in a single beam are transmitted by a 50:50 beam splitter can obtained
by taking the square root of eq.~(\ref{s2}),
\begin{equation}
\abs{A(k,n - k|n,0)} = \sqrt{C^{n}_{k}} \left( {1 \over 2} \right)^{n / 2}.
\label{s6}
\end{equation}
These amplitudes have the obvious symmetries,
\begin{equation}
\abs{A(k,n - k|n,0)} = \abs{A(n - k,k|n,0)} = \abs{A(k,n - k|n,0)} = \abs{A(n - k,k|n,0)}.
\label{s7}
\end{equation}
We must also consider the phases of these amplitudes, or at least the relative phases.
The hint is that we may consider the phase of a reflected photon to be shifted with
respect to that of a transmitted photon by $90^\circ$, as follows from a classical
analysis of waves in a 50:50 beam splitter \cite{set4} (see also the Appendix).  
In this problem, we
define the phase of a transmitted photon to be zero, so that the probability
amplitude should include a factor of $i = \sqrt{-1}$ for each reflected photon.
Thus, we have
\begin{eqnarray}
A(k,n - k|n,0) & = & i^{n - k} \sqrt{C^{n}_{k}} \left( {1 \over 2} \right)^{n / 2},
\label{s8} \\
A(n - k,k|n,0) & = & i^{k} \sqrt{C^{n}_{k}} \left( {1 \over 2} \right)^{n / 2},
\label{s9} \\
A(k,n - k|n,0) & = & i^{k} \sqrt{C^{n}_{k}} \left( {1 \over 2} \right)^{n / 2},
\label{s10} \\
A(n - k,k|n,0) & = & i^{n - k} \sqrt{C^{n}_{k}} \left( {1 \over 2} \right)^{n / 2}.
\label{s11}
\end{eqnarray}

\subsection{Two Input Beams}

We now calculate the general probability $P(N_1,n_1 + n_2 - N_1|n_1,n_2)$ that
$N_1$ output photons are observed along the direction of input beam 1 when
the number of photons in the input beams in $n_1$ and $n_2$.

We first give a classical wave analysis.  The input waves have amplitudes
$a_{1,2} = \sqrt{n_{1,2} \hbar \omega}$, and are in phase at the center of
the beam splitter.  The output amplitudes are the sums of the reflected and
transmitted parts of the input amplitudes.  A relected amplitude has a phase
shift of $90^\circ$ relative to its corresponding transmitted amplitude, as
discussed in sec.~2.1.  In the 50:50 beam splitter, the magnitude of both
the reflected and transmitted amplitudes from a single input beam are $1 / \sqrt{2}$
times the magnitude of the amplitude of that beam.
Hence, the output amplitudes are
\begin{eqnarray}
A_1 & = & {1 \over \sqrt{2}} (a_1 + i a_2),
\label{s12} \\
A_2 & = & {1 \over \sqrt{2}} (i a_1 + a_2).
\label{s13}
\end{eqnarray}
Taking the absolute square of eqs.~(\ref{s12})-(\ref{s13}), we find the output
beams to be described by
\begin{equation}
N_{1,2} = {\abs{A_{1,2}}^2 \over \hbar \omega} =  {a_1^2 + a_2^2 \over 2 \hbar \omega}
= {n_1 + n_2 \over 2}\, .
\label{s14}
\end{equation}
The classical view is that a 50:50 beam splitter simply splits both input beams, when
they are in phase.

For a quantum analysis,
we proceed by noting that of the $N_1$ photons in output beam 1, $k$ of these
could have come by transmission from input beam 1, and $N_1 - k$ by reflection
from input beam 2 (so long as $N_1 - k \leq n_2$).  The probability amplitude
that $k$ out of $N_1$ photons are transmitted from beam 1 while $N_1 - k$
photons are reflected from beam 2 is, to within a phase  factor, the
product of the amplitudes for each of these configurations resulting from
a single input beam:
\begin{equation}
A(k,N_1 - k|n_1,0) A(N_1 - k,n_2 - N_1 + k|0,n_2)
= (-1)^{n_1 - k} \sqrt{C^{n_1}_{k} C^{n_2}_{N_1 -k}} \left( {1 \over 2} \right)^{(n_1 + n_2) / 2},
\label{s22}
\end{equation}
referring to eqs.~(\ref{s8})-(\ref{s11}).  The most dramatic nonclassical features
to be found below can be attributed to the presence of the factor $(-1)^{n_1 - k}$
that arises from the $90^\circ$ phase shift between reflected and transmitted photons.

Since photons obey Bose statistics, we sum the sub-amplitudes (\ref{s22}),
weighting each one by the square root of the number of ways that $k$ out of the $N_1$ photons in
the first output beam can be assigned to input beam 1, namely $C^{N_1}_k$, time
the square root of the number of ways that the remaining $n_1 - k$ photons from
input beam 1 can be assigned to the $N_2$ photons in output beam 2, namely $C^{N_2}_{n_1 - k}$
 to obtain\footnote{Delicate to justify not also including factors $C^{N_1}_{N_1 - k}$, and
$C^{N_2}_{N_2 - (n_1 - k)}$, these being the ways of assigning photons to output beam 2 -- but these
factors are the same as those already included, and so should not be counted twice...}
\begin{eqnarray}
A(N_1,n_1 + n_2 - N_1|n_1,n_2)
& = & \sum_k \sqrt{C^{N_1}_{k} C^{N_2}_{n_1 -k}} A(k,N_1 - k|n_1,0) A(N_1 - k,n_2 - N_1 + k|0,n_2)
\nonumber \\
& = & (-1)^{n_1} \left( {1 \over 2} \right)^{(n_1 + n_2) / 2} 
\sum_k (-1)^{k} \sqrt{C^{n_1}_{k} C^{n_2}_{N_1 -k} C^{N_1}_{k} C^{N_2}_{n_1 -k}}.
\label{s23}
\end{eqnarray}
When evaluating this expression, any  binomial coefficient $C^n_m$ in which $m$ is negative, or
greater than $n$, should be set to zero.

The desired probability is, of course,
\begin{equation}
P(N_1,n_1 + n_2 - N_1|n_1,n_2) = \abs{A(N_1,n_1 + n_2 - N_1|n_1,n_2)}^2
\label{s24}
\end{equation}

Some examples of the probability distributions for small numbers of input photons are
given below.

\subsubsection{Two Input Photons}

\begin{center}
\begin{tabular}{c|ccc}
 Input & \multicolumn{3}{c}{Output $(N_1,N_2|$} \\
$|n_1,n_2)$ & $(0,2|$ & $(1,1|$ & $(2,0|$ \\
\hline
$|2,0)$ & ${1 \over 4}$ & ${1 \over 2}$ & ${1 \over 4}$ \\
$|1,1)$ & ${1 \over 2}$ & ${0}$ & ${1 \over 2}$ \\
$|0,2)$ & ${1 \over 4}$ & ${1 \over 2}$ & ${1 \over 4}$ \\
\end{tabular}
\end{center}

When $n_1$ or $n_2$ is zero, the probability distribution is binomial, as found
in sec.~2.1.  When $n_1 = n_2 = 1$ there is complete destructive interference
between the cases where both photons are reflected (combined phase shift = $180^\circ$)
and when both are transmitted (combined phase shift = 0).  This quantum result is
strikingly different from the classical expectation that there would be one photon
in each output beam.

\subsubsection{Three Input Photons}

\begin{center}
\begin{tabular}{c|cccc}
Input & \multicolumn{4}{c}{Output $(N_1,N_2|$} \\
$|n_1,n_2)$ & $(0,3|$ & $(1,2|$ & $(2,1|$ & $(3,0|$ \\
\hline
$|3,0)$ & ${1 \over 8}$ & ${3 \over 8}$ & ${3 \over 8}$ & ${1 \over 8}$ \\
$|2,1)$ & ${3 \over 8}$ & ${1 \over 8}$ & ${1 \over 8}$ & ${1 \over 8}$ \\
$|1,2)$ & ${3 \over 8}$ & ${1 \over 8}$ & ${1 \over 8}$ & ${1 \over 8}$ \\
$|0,3)$ & ${1 \over 8}$ & ${3 \over 8}$ & ${3 \over 8}$ & ${1 \over 8}$ \\
\end{tabular}
\end{center}

\subsubsection{Four Input Photons}

\begin{center}
\begin{tabular}{c|ccccc}
Input & \multicolumn{5}{c}{Output $(N_1,N_2|$} \\
$|n_1,n_2)$ & $(0,4|$ & $(1,4|$ & $(2,2|$ & $(3,1|$ & $(4,0|$ \\
\hline
$|4,0)$ & ${1 \over 16}$ & ${1 \over 4}$ & ${3 \over 8}$ & ${1 \over 4}$ & ${1 \over 16}$ \\
$|3,1)$ & ${1 \over 4}$ & ${1 \over 4}$ & ${0}$ & ${1 \over 4}$ & ${1 \over 4}$ \\
$|2,2)$ & ${3 \over 8}$ & ${0}$ & ${1 \over 4}$ & ${0}$ & ${3 \over 8}$ \\
$|1,3)$ & ${1 \over 4}$ & ${1 \over 4}$ & ${0}$ & ${1 \over 4}$ & ${1 \over 4}$ \\
$|0,4)$ & ${1 \over 16}$ & ${1 \over 4}$ & ${3 \over 8}$ & ${1 \over 4}$ & ${1 \over 16}$ \\
\end{tabular}
\end{center}

\subsubsection{Symmetric Input Beams: $n_1 = n_2 \equiv n$}

There is zero probability of observing an odd number of photons in either
output beam.

To see this, note that when $n_1 = n_2 = n$, the magnitudes of the subamplitudes
are equal for having $k$ photons appearing in output beam 1 from either 
input beam 1 or input beam 2.
However, the phases of these two subamplitudes are $180^\circ$ apart, so that
they cancel.  In particular, when $k$ photons are transmitted into output beam 1
from input beam 1, then $N_1 - k$ photons are reflected from input beam 2 into
output beam 1; meanwhile, $n - k$ photons are reflected from input beam 1 into
ouptput beam 2.  So the overall phase factor of this subamplitude is
$i^{N_1 - k + n - k} = (-1)^k\ i^{n + N_1}$.
Whereas, if $k$ photons are reflected from input beam 2 into output beam 1,
then $N_1 - k$ photons are transmitted from input beam 1 into output beam 1,
and so $n - N_1 + k$ photons are reflected from input beam 1 into output
beam 2.  So the overall phase factor of this subamplitude is
$i^{k + n - N_1 + k} = (-1)^k\ i^{n - N_1}$.  The phase factor between
these two subamplitudes (whose magnitudes are equal) is $i^{2 N_1} = (-1)^{N_1}$, which
is $-1$ for odd $N_1$, as claimed.

For the case of observing an even number of photons in the output beams, a
remarkable simplification of eq.~(\ref{s23}) holds \cite{Campos}. {\sl I have not been able to
show this by elementary means.  It does follow by inspection when $m = 0$ or $n$, in which case
eq.~(\ref{s23}) contains only a single nonzero term.  In general, the index $k$ in eq.~(\ref{s23})
for $A(2m,2n-2m|n,n)$ runs from 0 to $2m$ if $2 m \leq n$,
or from $2m - n$ to $n$ if $2m \geq n$.
There are an odd number of terms, the central one having index $k = m$.  By a
strange miracle of combinatorics, the sum collapses to a simplified version of the central
term of the series....}   Namely,
\begin{equation}
A(2m,2n-2m|n,n) = (-1)^{n-m} \left( {1 \over 2} \right)^{n} \sqrt{C^{2m}_{m} C^{2n-2m}_{n-m}}.
\label{s25}
\end{equation}
Therefore, the $n + 1$ nonvanishing probabilities for symmetric input beams are
\begin{equation}
P(2m,2n-2m|n,n) =  \left( {1 \over 2} \right)^{2n} C^{2m}_{m} C^{2n-2m}_{n-m}
\approx {1 \over n \pi \sqrt{{m \over n}(1 - {m \over n})}}\, ,
\label{s26}
\end{equation}
where the approximation holds for large $m$ and large $n$.
Note that $\int_0^1 dx / \sqrt{x (1 - x)} = \pi$.
This probability distribution peaks for $m = 0$ or $n$, \ie, for all photons in one
or the other output beam, with value
\begin{equation}
P(0,2n|n,n) =  P(2n,0 |n,n) =\left( {1 \over 2} \right)^{2n} C^{2n}_{n}.
\label{s27}
\end{equation}
The probability of finding all output photons in a single beam when the input beams
are symmetric is larger by a factor $C^{2n}_{n}$ than when there is only a single
input beam (of the same total number of photons), because there are $C^{2n}_{n}$
ways of assigning the $n$ photons from input beam 1 to the $2n$ photons in the
output beam.  This is an extreme example of
photon bunching caused by the beam splitter.

{\sl However, I reamin uncomfortable with the result (\ref{s26}) because it does not
agree with the classical prediction (\ref{s14}) in the large $n$ limit.....}

\section{Appendix: Phase Shift in a Lossless Beam Splitter}

We give a classical argument based on a Mach-Zender interferomter, shown in the figure below,
that there is a $90^\circ$ phase shift between the reflected and transmitted beams
in a lossless beam splitter.  Then, following Dirac's dictum, we suppose that this
result applies to a single photon.

A beam of light of unit amplitude is incident on the interferometer from the upper
left.  The reflected and transmitted amplitudes are $r e^{i \phi_r}$ and $t e^{i \phi_t}$,
where magnitudes $r$ and $t$ are real numbers.  The condition of a lossless beam 
splitter is that
\begin{equation}
r^2 + t^2 = 1.
\label{a1}
\end{equation}
The reflected and transmitted beams are reflected off mirrors and recombined in a 
second lossless beam splitter, identical to the first.  

\medskip
\centerline{\includegraphics*[width=5in]{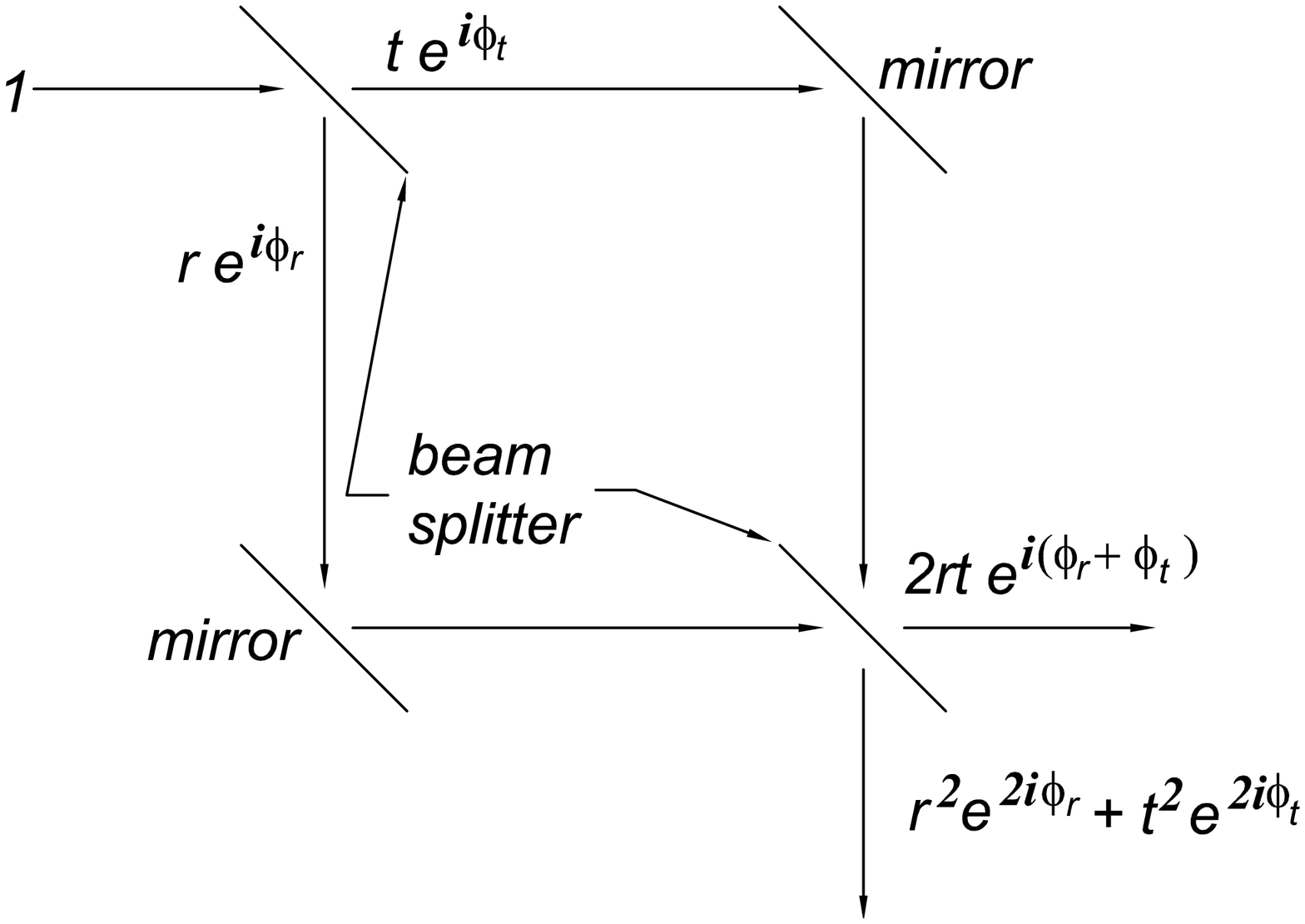}}

Then, the amplitude for transmission at the first beam splitter, followed by
reflection at the second, is $tr e^{i(\phi_t + \phi_r)}$, \etc\ \ 
Hence, the recombined beam that moves to the right has amplitude
\begin{equation}
A_1 = 2 rt e^{i(\phi_r + \phi_t)},
\label{a2}
\end{equation}
while the recombined beam that moves downwards has amplitude
\begin{equation}
A_2 = r^2 e^{2i \phi_r} + t^2 e^{2i \phi_t}.
\label{a3}
\end{equation}
The intensity of the first output beam is
\begin{equation}
I_1 = \abs{A_1}^2 = 4 r^2 t^2,
\label{a4}
\end{equation}
and that of the second output beam is
\begin{equation}
I_2 = \abs{A_2}^2 = r^4 + t^4 + 2 r^2 t^2 \cos 2(\phi_t - \phi_r).
\label{a5}
\end{equation}
For lossless splitters, the total output intensity must be unity,
\begin{equation}
I_1 + I_2 = 1 = (r^2 + t^2)^2 + 2 r^2 t^2 [ 1 + \cos 2(\phi_t - \phi_r)].
\label{a6}
\end{equation}
Recalling eq.~(\ref{a1}), we must have
\begin{equation}
\phi_t - \phi_r = \pm 90^\circ,
\label{a7}
\end{equation}
for any value of the splitting ratio $r^2:t^2$.


\begin{thebibliography}{99}

\bibitem{Dirac}
P.A.M.~Dirac,
{\em The Principles of Quantum Mechanics}, 4th ed.\
(Clarendon Press, London, 1958), p.~9.

\bibitem{set4}
See, for example, Problem 4 of the Princeton Ph501 problem set at \hfill\break
http://puhep1.princeton.edu/\~mcdonald/examples/ph501set6.pdf \hfill\break
Early discussions of phase shifts in beam splitters include 
G.B.~Airy,
Phil.\ Mag.\ {\bf 2}, 20 (1833);
G.G.~Stokes,
Cambridge and Dublin Math.\ J.\ {\bf 4}, 1 (1849), reprinted in
{\em Mathematical and Physical Papers of G.G.~Stokes}, V.~2
(Cambridge U.\ Press, Cambridge, 1883), pp.~89-103.

\bibitem{Campos}
R.A.~Campos, B.E.A.~Saleh and M.C.~Teich,
{\sl Quantum-mechanical lossless beam splitter: SU(2) symmetry and photon
statistics},
Phys.\ Rev.\ A {\bf 40}, 1371 (1989).

\bibitem{Paris}
M.G.A.~Paris,
{\sl Homodyne Photocurrent, Symmetries in Photon Mixing and Number
State Synthesis},
Int.\ J.\ Mod.\ Phys.\ B {\bf 11}, 1913 (1997).

\bibitem{Feynman}
R.P.~Feynman, R.B.~Leighton and M.~Sands,
{\em The Feynman Lectures on Physics}
(Addison Wesley, Reading, MA, 1965), Vol.~III, secs.~4.2-3.

\bibitem{Hong}
C.K.~Hong, Z.Y.~Ou and L.~Mandel,
{\sl Measurement of Subpicosecond Time Intervals between Two Photons
by Interference},
Phys.\ Rev.\ Lett.\ {\bf 59}, 2044 (1987).

\bibitem{Ou}
Z.Y.~Ou, J.-K.~Rhee and L.J.~Wang,
{\sl Observation of Four-Photon Interference with a Beam Splitter
by Pulsed Parametric Down-Conversion},
Phys.\ Rev.\ Lett.\ {\bf 83}, 959 (1999);
{\sl Photon bunching and multiphoton interference in parametric down-conversion},
Phys.\ Rev.\ A {\bf 60}, 593 (1999).

\end{thebibliography}
\end{document}